\magnification\magstep1



\hbadness=10000      
\vbadness=10000  

\font\eightit=cmti8  
\font\eightrm=cmr8 \font\eighti=cmmi8                 
\font\eightsy=cmsy8 
           \font\sixrm=cmr6



\def\eightpoint{\normalbaselineskip=10pt 
\def\rm{\eightrm\fam0} \let\it\eightit
\textfont0=\eightrm \scriptfont0=\sixrm 
\textfont1=\eighti \scriptfont1=\seveni
\textfont2=\eightsy \scriptfont2=\sevensy 
\normalbaselines \eightrm
\parindent=1em}



\def\eq#1{{\noexpand\rm(#1)}}          
\newcount\eqcounter                    
\eqcounter=0                           
\def\numeq{\global\advance\eqcounter by 1\eq{\the\eqcounter}}           
\def\relativeq#1{{\advance\eqcounter by #1\eq{\the\eqcounter}}}


\def\namelasteq#1{\global\edef#1{{\eq{\the\eqcounter}}}}  


\def\cite#1{{\rm[#1]}}                 

\def\ren{{\rm ren}}


\newif\ifstartsec                   

\outer\def\section#1{\vskip 0pt plus .15\vsize \penalty -250
\vskip 0pt plus -.15\vsize \bigskip \startsectrue
\message{#1}\centerline{\bf#1}\nobreak\noindent}

\def\subsection#1{\ifstartsec\medskip\else\bigskip\fi \startsecfalse
\noindent{\it#1}\penalty100\medskip}

\def\refno#1. #2\par{\smallskip\item{\rm\lbrack#1\rbrack}#2\par}

\hyphenation{geo-me-try}


\def\Doplicher{1}
\def\CDS{2}
\def\Mtheory{3}
\def\KLM{4}
\def\Madore{5}
\def\Grosse{6}
\def\Filk{7}
\def\VarPP{8}
\def\Chaichian{9}
\def\Instant{10}
\def\Book{11}
\def\Landi{12}
\def\Sorella{13}
\def\Breitenlohner{14}
\def\Speer{15}
\def\KW{16}


\rightline{FT/UCM--15--99}

\vskip 1cm

\centerline{\bf The One-loop UV Divergent Structure of $U(1)$ Yang-Mills 
Theory}
\centerline{\bf on  Noncommutative ${\rm{I\!R}}^4$}

\bigskip

\centerline{\rm C. P. Mart{\'\i}n* and D. S\'anchez-Ruiz\dag}
\medskip
\centerline{\eightit Departamento de F{\'\i}sica Te\'orica I,						Universidad Complutense, 28040 Madrid, Spain}
\vfootnote*{email: {\tt carmelo@elbereth.fis.ucm.es}}
\vfootnote\dag{email: {\tt domingos@eucmos.sim.ucm.es}}

\bigskip\bigskip

\begingroup\narrower\narrower
\eightpoint
We show that $U(1)$ Yang-Mills theory on noncommutative ${\rm {I\!R}}^4$ 
can be renormalized at the one-loop level by multiplicative dimensional 
renormalization of the coupling constant and fields of the theory. 
We compute the beta function of the theory and conclude that the theory is
asymptotically free. We also show that the Weyl-Moyal matrix defining
the deformed product over the space of functions on ${\rm {I\!R}}^4$ is not
renormalized at the one-loop level.  

\par
\endgroup 
\vskip 2cm
Field theories on noncommutative spaces may play an important role in 
unraveling the properties of Nature at the Planck scale 
\cite{\Doplicher,\CDS}. Yang-Mills theories on  the noncommutative torus
occur in compactifications of M-theory \cite{\CDS} and there is already
a good many papers where M-theory  on noncommutative tori has been studied
\cite{\Mtheory}. Still, it has not been established yet that these field
theories are well-defined at the quantum level. Although a few
initial steps were taken in ref. \cite{\Doplicher} (see also \cite{\KLM}) 
the very definition of quantum field theory over 
noncommutative spaces (the analog of the
standard Wightman-Osterwalder-Schrader axioms, scattering theory, ...) 
is yet to be stated. 
Quantum Field theories on  fuzzy spheres \cite{\Madore} had been studied 
previously and its UV finiteness established \cite{\Grosse}. Here the
number of quantum degrees of freedom is finite and hence it seems that 
the quantum theory exists. This is not clear at all for field theories over 
noncommutative spaces such as
the noncommutative ${\rm {I\!R}}^4$ \cite{\Filk}, the noncommutative
3-tori \cite{\VarPP} or the noncommutative plane \cite{\Chaichian}.
In these cases the quantum field theory has UV divergences and there 
remains the question whether this theories are renormalizable. 
It should be noticed that for the theories at hand  the 
interaction  terms in Fourier space are no longer polynomials in the 
momenta, and hence  that the renormalization program (either perturbative or 
non-perturbative) works cannot be taken for granted. 

 The purpose of this article is to analyze the one-loop UV divergent structure
of the simplest pure gauge theory over the noncommutative 
${\rm {I\!R}}^4$. This theory is a $U(1)$ theory but it is not a free
theory due to the noncommutative character of the base space. Indeed,
now the field strength (curvature) is no longer linear in the gauge field 
(connection), but it contains an antisymmetric quadratic contribution as well,
for the product of functions is commutative no longer. The theory 
with classical action given by the Yang-Mills functional turns out to be an 
interacting theory.

The noncommutative ${\rm {I\!R}}^4$ is defined as the algebra generated
by four hermitian elements $x_{\mu}$, $\mu=1,2,3,4$, which verify the equation
$$
[x_\mu,x_\nu]=iQ_{\mu\nu},
$$ 
where $Q_{\mu\nu}$ is a real constant antisymmetric matrix of rank 4. 
Equations over the noncommutative ${\rm {I\!R}}^4$ can be represented 
as equations over a deformation of the $C^{*}$ algebra 
$C_{\infty}({\rm {I\!R}}^4)$ of continuous complex-valued functions 
over ${\rm {I\!R}}^4$ vanishing at infinity  \cite{\Instant}. 
This deformation is given by the the Weyl product
$$
(f\star g)(x)=\int\int {d^4 q\over (2\pi)^4} {d^4 p\over (2\pi)^4}\;
e^{i\,\omega_{\mu\nu}q_{\mu}p_{\nu}}\;f(q)g(p).
$$
Here $f(q)$ and $q(p)$ are, respectively, the Fourier transforms of 
$f(x)$ and $g(x)$, and $\omega_{\mu\nu}\equiv{1\over 2}(Q^{-1})_{\mu\nu}$
is a constant antisymmetric matrix of rank four. 
Connes's noncommutative geometry formalism gives mathematical rigour 
to the concept of classical $U(1)$ gauge field over such a noncommutative 
space \cite{\Book,\Landi}. This gauge field is provided by a real vector
function, $A_{\mu}(x)$, on ${\rm {I\!R}}^4$. The field strength, 
$F_{\mu\nu}(x)$, for this gauge field reads now 
$$
F_{\mu\nu}(x)=\partial_{\mu}A_{\mu}-\partial_{\nu}A_{\mu}+i
\{A_{\mu},A_{\nu}\}(x),
$$ 
where $\{A_{\mu},A_{\nu}\}(x)=
(A_{\mu}\star A_{\nu})(x)-(A_{\nu}\star A_{\mu})(x)$ is 
the Moyal bracket. 

The classical $U(1)$ field theory over noncommutative ${\rm {I\!R}}^4$ is given
by the Yang-Mills functional
$$ 
 S_{YM}={1\over4 g^2}\int( F_{\mu\nu}\star F_{\mu\nu}) (x),\eqno\numeq
$$\namelasteq\YMaction
where $F_{\mu\nu}(x)$ is given above. This action is invariant under 
gauge transformations which have the following infinitesimal form
$\delta A_{\mu}(x)= \partial_{\mu}\theta+i\{A_{\mu},\theta\}(x)$. 

The next task to tackle is the construction of a quantum field theory 
whose classical counterpart is the previous theory. What we mean by a 
quantum field theory over a noncommutative space is by no means obvious, 
{\it e.g.} what mathematical objects define the quantum physics has not 
been properly established yet (see ref. \cite{\Chaichian} for discussions 
on this point). In this paper we shall assume that the quantum theory is
defined by the Green functions of the theory, {\it i.e.} by the
generating functional
$$
Z[j]=N\int {\cal D}\phi(x) e^{-S+\int d^4 x j(x)\phi(x)}\equiv
N e^{-S_{\rm int}[{\delta{\,\,}\over\delta J(x)}]}
{\rm exp}\big({1\over 2}\int\!\!\int d^4x d^4y j(x)P(x-y)j(y)\big),
$$ 
where $\phi$ denotes generically the ``fields'' of the theory and 
$P(x-y)$ denotes the inverse of the kinetic term (quadratic in the 
``fields'') in $S$. $S$ and $S_{\rm int}$ denote, respectively, the classical 
action over the noncommutative space in question and the interaction terms in
in $S$. The previous definition 
of $Z[j]$ is to be understood as a formal expansion in terms of Feynman 
diagrams. 

Since our Yang-Mills action is invariant under  gauge transformations its
kinetic term has no inverse and a gauge-fixing term has to be introduced.
We shall do this in a consistent way by using the BRS formalism.
Let us introduce the ghost ``fields'' $c$ and $\bar {c}$, the gauge-fixing
field $B$ and define the BRS transformations as follows
$$
sA_{\mu}(x)=D_{\mu}c(x)=\partial_{\mu}c(x)+i\{A_{\mu},c\}(x), 
s{\bar c}(x)=b, sb(x)=0, sc(x)=-(c\star c)(x). 
$$
To keep track of the renormalization of the composite transformations 
$sA_{\mu}(x)$ and $sc(x)$ one also introduces the external fields
$J_{\mu}(x)$ and $H(x)$ which couple to them \cite{\Sorella}. 
The BRS invariant classical  $4$-dimensional action is
$$
S_{\rm cl} = S_{\rm YM} + S_{\rm gf} + S_{\rm ext},      \eqno\numeq
$$\namelasteq\BRSaction
where $S_{YM}$ has been given in eq.~\YMaction\ and
$$
\eqalignno{
S_{\rm gf} &= \int d^4x \; s[{\bar c}\star({\alpha\over 2} B-
\partial_\mu A^\mu)](x),\cr
S_{\rm ext} &= \int d^4x \; 
\Bigl(J^\mu\star sA_\mu + H\star sc\Bigr)(x).        \cr
}
$$
 Now, standard path integral formal manipulations lead to the Slavnov-Taylor
identity for the 1PI functional $\Gamma[A_\mu,B,c,{\bar c};J_\mu, H]$.
This identity reads
$$
{\cal S}(\Gamma)\equiv
\int d^4 x\,\,\Bigl[
 \,{\delta\Gamma\over\delta J^\mu}
 {\delta\Gamma\over\delta A_\mu}+
 \,{\delta\Gamma\over\delta H}
 {\delta\Gamma\over\delta c}+
 \; B {\delta\Gamma\over\delta\bar c}\Bigr]=0. \eqno\numeq
$$\namelasteq\STidentity
The Slavnov-Taylor identity governs the BRS symmetry of the theory 
at the quantum level.
 
Since the formal Feynman diagrams contributing to 
$\Gamma[A_\mu,B,c,{\bar c};J_\mu, H]$ present UV divergences it is not 
straightforward that the renormalized, would it exist, 1PI functional 
defining the quantum theory satisfies the Slavnov-Taylor identity: anomalies
may occur. We shall see in this paper that as far as our explicit
computations reach no anomalies occur in the quantum theory and that indeed
a renormalized BRS invariant 1PI functional does exist. We shall be 
working only at the one-loop level. 

First, it is not difficult to see that one-loop UV divergent 1PI Green 
functions are following: $\Gamma_{AA}$,  $\Gamma_{AAA}$, $\Gamma_{AAAA}$,
$\Gamma_{{\bar c} c}$,  $\Gamma_{{\bar c}A c}$, $\Gamma_{J c}$,
$\Gamma_{J A c}$ and $\Gamma_{Hcc}$, with obvious notation. Notice that the
Feynman rules from eq.~\BRSaction\ are obtained from the Feynman rules
for  $SU(N)$ Yang-Mills theory on commutative Euclidean space upon
the replacement $f_{a_1 a_2 a_3}\rightarrow 2\sin \omega(p_2,p_3)$, 
$p_2$ and $p_3$ being respectively the
momenta carried by the lines with colour index $a_2$ and $a_3$. 
Hence, the counterpart in our $U(1)$ theory of a Feynman diagram which is 
finite by power-counting in the standard $SU(N)$ theory will also be finite.
Since no Action Principle (see ref. \cite{\Sorella} and references 
therein) has been shown to hold yet for field theories over noncommutative
spaces, we cannot carry out a cohomological study of the renormalizability
of the theory at hand. We shall proceed by performing explicit computations.

To regularize the Feynman integrals of our theory will shall use dimensional
regularization. The dimensionally regularized counterpart of any Feynman
diagram is  defined as follows: first, $2i\sin \omega(p_2,p_3)$ is
expressed as $e^{i \omega(p_2,p_3)}-e^{-i \omega(p_2,p_3)}$; second,  the
four-dimensional momentum measure ${d^4 p\over (2\pi)^4}$ is replaced with the
``$D$-dimensional'' measure ${d^D p\over (2\pi)^D}$ and any 4-dimensional 
algebraic expression with a ``$D$-dimensional'' one defined 
according to the rules in ref. \cite{\Breitenlohner}; third, 
gaussian integration over the ``$D$-dimensional''  loop momenta is 
carried out, which leads to an integral over the  $\alpha$-parameter  space of 
Schwinger introduced by using the following equation \cite{\Speer} 
$$
                 {1\over (p_i^2)^m}=
{1\over \Gamma(m)}\int_{0}^{\infty}d\alpha_i\, 
\alpha_i^{m-1}\,e^{-\alpha_i p_i^2};
$$
and, four,  $D$ is promoted to a complex variable and any formal tensor 
expression is defined to be an algebraic expression satisfying only the 
algebraic rules in ref.\cite{\Breitenlohner}. In the theory under study we
have a new algebraic object to be defined in space of ``$D$-dimensional'' 
algebraic objects. This is the antisymmetric matrix $\omega_{\mu\nu}$, which
has rank four, {\it i.e.} $\omega_{\mu\nu}=-\omega_{\nu\mu}$ and 
$(\omega_{\mu\nu}p_{\nu})^2>0$ if $p\not= 0$. These properties are only
compatible for even-dimensional spaces, hence, we shall define the
``$D$-dimensional'' algebraic object $\omega_{\mu\nu}$ as being 
``intrinsically 4-dimensional'' and having the aforementioned properties:
$$\
\omega_{\mu\nu}{\hat g}_{\nu\rho} =0,\quad
\omega_{\mu\nu}=-\omega_{\nu\mu},\quad
\omega_{\rho\mu}\omega_{\rho\nu}{\bar p}_{\mu}{\bar p}_{\nu}>0\; 
{\rm if}\; {\bar p}_{\mu}\not= 0,\quad 
{\bar p}_{\mu}\equiv {\bar g_{\mu\nu}}p_{\nu}.
$$
The symbols ${\bar g}_{\mu\nu}$,${\hat g}_{\nu\rho}$ denote respectively 
the ``4-dimensional'' and ``$(D-4)$-dimensional'' metrics as
defined in ref.  \cite{\Breitenlohner}.
With these definitions it is not difficult to show that at the one-loop level,
and provided we keep away from exceptional external momenta, the UV divergences
occur as simple poles at $D=4$ whose residue is a sum of monomials,
each monomial being the product of a polynomial on the external momenta
(with no ``hatted'' metric ${\hat g}_{\nu\rho}$) and a complex exponential 
whose argument is as linear combination of terms
of the type $\omega_{\mu\nu}p^{\mu}_{i}p^{\nu}_{j}$ ($p_i$ and $p_j$ denote
external momenta). Indeed, every time a exponential carrying a loop momenta
occurs in the integrand, it turns out that its dimensionally regularized
counterpart is finite at $D=4$. Hence, if BRS invariance is not broken
it would be likely that one-loop renormalization can be achieved by 
multiplicative renormalization of the parameters of the theory.

We have computed the one-loop UV divergent contribution to all the divergent
1PI Green functions. The results we have obtained read thus
$$
\eqalignno{
\Gamma_{\mu_1\mu_2}^{(AA)}(p)=&-\Bigl({1\over(4\pi)^2\varepsilon}\Bigr)
   \Bigl({13\over 3} -\alpha \Bigr)
       \Bigl(p_{\mu_1}p_{\mu_2}-p^2\,g_{\mu_1\mu_2}\Bigr)\,\cr
\Gamma_{\mu_1\mu_2\mu_3}^{(AAA)}(p_1,p_2,p_3)=&-
i\Bigl({1\over(4\pi)^2\varepsilon}\Bigr)
\Bigl({17\over 3} -3\alpha \Bigr)\;
             \sin[\omega_{\mu\nu}(p_2)_{\mu}(p_3)_{\nu}]\cr
&\Bigl((p_1-p_3)_{\mu_2}g_{\mu_1\mu_3}+(p_2-p_1)_{\mu_3}g_{\mu_1\mu_2}+
(p_3-p_2)_{\mu_1}g_{\mu_2\mu_3}\Bigr)\,\cr
\Gamma_{\mu_1\mu_2\mu_3\mu_4}^{(AAAA)}(p_1,p_2,p_3,p_4)=&
\Bigl({1\over(4\pi)^2\varepsilon}\Bigr)
\Bigl({ 4\over 3} - 2\alpha \Bigr)\cr
4\bigl[&\sin[\omega_{\mu\nu}(p_1)_{\mu}(p_2)_{\nu}]
\sin[\omega_{\mu\nu}(p_3)_{\mu}(p_4)_{\nu}]
(g_{\mu_1\mu_3}g_{\mu_2\mu_4}-g_{\mu_1\mu_4}g_{\mu_2\mu_3})+\cr
&\sin[\omega_{\mu\nu}(p_1)_{\mu}(p_4)_{\nu}]
\sin[\omega_{\mu\nu}(p_3)_{\mu}(p_2)_{\nu}]
(g_{\mu_2\mu_4}g_{\mu_1\mu_3}-g_{\mu_4\mu_3}g_{\mu_1\mu_2})+\cr
&\sin[\omega_{\mu\nu}(p_4)_{\mu}(p_2)_{\nu}]
\sin[\omega_{\mu\nu}(p_3)_{\mu}(p_1)_{\nu}]
(g_{\mu_4\mu_3}g_{\mu_2\mu_1}-g_{\mu_1\mu_4}g_{\mu_2\mu_3})\bigr]\cr
\Gamma^{({\bar c} c)}(p)=&-\Bigl({1\over(4\pi)^2\varepsilon}\Bigr)\;
\Bigl({1\over 2}\Bigr)\Bigl(3-\alpha\Bigr)\,g^2\;p^2 \cr
\Gamma_{\mu_2}^{({\bar c}A c)}(p_1,p_2,p_3)=&-i
\Bigl({1\over(4\pi)^2\varepsilon}\Bigr)
\bigl(\alpha g^2 \bigr)\;(p_1)_{\mu_2}\;
            2\, \sin[\omega_{\mu\nu}(p_2)_{\mu}(p_3)_{\nu}]\cr
\Gamma_{\mu}^{(Jc)}(p)=&\Bigl({1\over(4\pi)^2\varepsilon}\Bigr)
   \Bigl({3 -\alpha\over 2}\Bigr)\, g^2\; p_\mu \cr
\Gamma_{\mu_1\mu_2}^{(JAc)}(p_1,p_2,p_3)=
&\Bigl({1\over(4\pi)^2\varepsilon}\Bigr)\;\bigl( \alpha g^2\bigr)\,
g_{\mu_1\mu_2}\; 
2\, \sin[\omega_{\mu\nu}(p_2)_{\mu}(p_3)_{\nu}]\cr
\Gamma^{(Hcc)}(p_1,p_2,p_3)=
&-\Bigl({1\over(4\pi)^2\varepsilon}\Bigr)\;\bigl( \alpha g^2\bigr)\;
e^{i[\omega_{\mu\nu}(p_2)_{\mu}(p_3)_{\nu}]},&\numeq\cr
}\namelasteq\Pole
$$
where $D=4-2\varepsilon$. Notice that the momentum structure 
 of the previous contributions is the same
as corresponding term in the BRS action in eq.~\BRSaction, upon formal
generalization of this action to the ``$D$-dimensional'' space of 
dimensional regularization. So, one  would expect that these 
1PI contributions can be subtracted by MS multiplicative renormalization 
of the fields and parameters in the BRS invariant action. 
And, indeed, this is so, if we perform the following infinite 
renormalizations  
$$
\eqalign{&g_0={\mu}^{2\varepsilon}\,Z_g\,g,\;\alpha_0=\,Z_{\alpha}\,\alpha,\;
A_{0\,\mu}=\,Z_A\,A_{\mu},\;B_0=\,Z_{B}\,B,\;\cr
&J_{0\,\mu}=\,Z_J\,J_{\mu},\;H_0=\,Z_{H}\,H,\;
c_0=\,Z_{c}\,c \quad{\rm and}\quad{\bar c}_0=Z_{\bar c}\,{\bar c},\cr
}
$$
where the subscript $0$ labels the bare quantities and
$$
\eqalignno{
&Z_g=1-{1\over(4\pi)^2\varepsilon}\,{22\over 3}\,g^2,\;
Z_A=1-{1\over(4\pi)^2\varepsilon}\,{3+\alpha\over 2}\,g^2\cr
&Z_{\bar c}Z_c=1+{1\over(4\pi)^2\varepsilon}\,{3-\alpha\over 2}\,g^2,\;
Z_{\bar c}Z_A Z_c=1-{1\over(4\pi)^2\varepsilon}\,\alpha\,g^2\cr
&Z_H Z_{c}^2=1-{1\over(4\pi)^2\varepsilon}\,\alpha\,g^2,
Z_B=Z_A^{-1},\;Z_{\alpha}=Z_A^2\quad {\rm and}\quad Z_J=Z_{\bar c}.&\numeq\cr
}\namelasteq\Zeds
$$
Notice that there is no renormalization of the  matrix $\omega_{\mu\nu}$.
That these $Z$s render UV finite the 1PI functions whose pole contribution
is in eq.~\Pole\ is a consequence of BRS invariance. Indeed, in view of
eq.~\Pole, it is not difficult to show that the singular contribution, 
$\Gamma^{({\rm pole})}$, to the dimensionally regularized 1PI functional can be
recast into the form
$$
\Gamma^{({\rm pole})}= 
{a\over 4g^2}\int d^Dx ( F_{\mu\nu}\star F_{\mu\nu}) (x) + 
{\cal B}_{D}\, X \eqno\numeq
$$\namelasteq\BRSpole
where 
$$
\eqalign{ 
&X=\int d^Dx \Bigl(a_1 
(J_{\mu}-\partial_{\mu}{\bar c})\star A_{\mu}-
a_2 H\star c\Bigr)(x),\cr
a=&{1\over(4\pi)^2\varepsilon}\,{22\over 3}\,g^2,\;
a_1=-{1\over(4\pi)^2\varepsilon}\,{3+\alpha\over 2}\,g^2,\;
a_2=-{1\over(4\pi)^2\varepsilon}\,\alpha\,g^2,\cr
}
$$
and ${\cal B}_{D}$ is the linearized Slavnov-Taylor operator acting upon the
space of formal algebraic expressions constructed with ``$D$-dimensional'' 
monomials of the fields and their derivatives. ${\cal B}_{D}$ is defined
as follows
$$
{\cal B}_{D}=
\int d^D x\,\,\Bigl[
 \,{\delta S_{\rm cl}\over\delta J^\mu}
 {\delta \over\delta A_\mu}+
  {\delta S_{\rm cl}\over\delta A^\mu}
 {\delta\over\delta J_\mu}+
 \,{\delta S_{\rm cl}\over\delta H}
 {\delta \over\delta c}+{\delta S_{\rm cl}\over\delta c}
 {\delta\over\delta H}+
 \; B {\delta\over\delta\bar c}\Bigr]. 
$$

The conclusion that one draws from eq.~\BRSpole\ is that the UV divergent
contributions displayed in eq.~\Pole\ are BRS invariant. Notice that 
$\Gamma^{({\rm pole})}$ is the sum of two terms: the second is 
${\cal B}_D$-exact (recall that ${\cal B}_D^2=0$), whereas the first, 
the Yang-Mills term, is ${\cal B}_D$-closed. This all goes hand in hand with
analysis of the UV divergent contributions in standard $SU(N)$ Yang-Mills 
theory. And, indeed, as  in standard four-dimensional Yang-Mills theory we have
$$ 
\eqalign{ &Z_g=1-a,\; Z_A=1+a_1,\; 
Z_{\bar c}Z_c=1-a_1+a_2,\;
Z_{\bar c}Z_A Z_c=1+a_2,\cr 
&Z_H Z_{c}^2=1+a_2,\;Z_B=Z_A^{-1},\;Z_{\alpha}=
Z_A^2\quad {\rm and}\quad Z_J=Z_{\bar c}.\cr
}
$$
Eq.~\Zeds\ is thus recovered. Let us remark that the values we have obtained
for the $Z$s agree with the corresponding values of the $Z$s of standard 
$SU(N)$ Yang-Mills theory on commutative ${\rm I\!R}^4$ upon replacing 
in the latter the constant $C_2(G)$ (the quadratic casimir in the adjoint 
representation) with $2$. Actually, the UV divergent contributions 
in eq.~\Pole\ agree with those in the standard $SU(N)$ Yang-Mills theory, 
if the following substitutions are made in the latter:
$f_{a_1 a_2 a_3}\rightarrow 2\sin \omega(p_2,p_3)$ and  
$C_2(G)\rightarrow 2$. Recall that the structure constants are not 
renormalized in standard $SU(N)$ Yang-Mills theory; 
$\omega_{\mu\nu}$, the matrix defining the Weyl-Moyal product, is 
not renormalized either.

We shall define the order $\hbar$ MS renormalized 1PI functional 
$\Gamma_{\ren}^{(1),\,{\rm MS}}$
as usual:
$$
\Gamma_{\ren}^{(1),\,{\rm MS}}=
{\cal LIM}_{\varepsilon\rightarrow 0}
\Bigl[\Gamma^{(1)}_{{\rm DReg}}-\Gamma^{({\rm pole})}\Bigr],
$$
where $\Gamma^{(1)}_{{\rm DReg}}$ denotes the dimensionally regularized
1PI functional at order $\hbar$ and  $\Gamma^{({\rm pole})}$ is given 
in eq.~\Pole. The limit $\varepsilon\rightarrow 0$ is taken after 
performing the subtraction of the pole  and replacing 
every ``$D$-dimensional'' algebraic object with its
4-dimensional counterpart \cite{\Breitenlohner}; this is 
 why we have denoted it by ${\cal LIM}$. 

Since  $\Gamma^{(1)}_{{\rm DReg}}$ is 
BRS invariant, {\it i.e} it satisfies the Slavnov-Taylor identity 
at order $\hbar$
$$
{\cal B}_{D}\Gamma^{(1)}_{{\rm DReg}}=0,
$$
the MS renormalized 1PI functional $\Gamma_{\ren}^{(1),\,{\rm MS}}$ is
also BRS invariant:
$$
{\cal B}\,\Gamma_{\ren}^{(1),\,{\rm MS}}=0.
$$
The operator ${\cal B}$ is the counterpart of ${\cal B}_{D}$ at $D=4$:
the linearized Slavnov-Taylor operator in noncommutative ${\rm I\!R}^4$.
We thus conclude that the Slavnov-Taylor identity (eq.~\STidentity)  holds 
for the renormalized theory at order $\hbar$. This statement is not
completely rigorous since there is no proof as yet that the Quantum Action
Principle \cite{\Breitenlohner} holds for the dimensionally regularized
amplitudes of the theory at hand. However, in our computations we have found 
no hint that this principle might not be valid here.

By using standard textbook techniques, one can work out the
renormalization group equation for $\Gamma_{\ren}^{\rm MS}$:
$$
\Bigl[\mu{\partial \over \partial\mu}-\beta{\partial \over \partial g}
-\delta_{\alpha}{\partial \over \partial \alpha}-\sum_{\phi}\gamma_{\phi}\,
\int d^4 x\,\phi(x){\delta \over \delta\phi(x)}\Bigr]\,
\Gamma^{\rm MS}_{\ren}[\phi;g,\omega,\alpha]=0.
$$
The fields are denoted  by $\phi$.    
It should be noticed
that $\omega_{\mu\nu}$ is a dimensionful parameter which does not run. 
The one-loop beta function of the theory is easily computed to be
$$
\beta(g^2)\equiv \mu{d g^2\over d\mu}= -{1\over 8\pi^2}\,{22\over 3}\,g^4.
$$
Hence, the theory is asymptotically free. The other renormalization
group coefficients read at the one-loop level:
$$
\eqalign{
&\gamma_{A}=+{1\over 8\pi^2}\big({3+\alpha\over 2}\bigr)\,g^4,\quad
\gamma_{c}=+{1\over 8\pi^2}\,\alpha\,g^4,\cr
&\gamma_{J}=\gamma_{\bar c}=\gamma_{B}=-\,\gamma_{A},\quad 
\delta_{\alpha}=-\,2\,\gamma_{A}\,\alpha,\quad\gamma_{H}=-\,\gamma_{c}.\cr
}
$$

We shall finish with two remarks. First, that the structure of the UV 
divergences, which is not a polynomial in momentum space, is a polynomial in
the fields and their derivatives with respect to the Weyl product. One
wonders whether this generalizes to higher loops upon subtraction of
subdivergences and whether the theory of normal products (on which 
the method of algebraic renormalization rests) remains valid upon replacing
the ordinary product with the Weyl product. Second, $\Gamma^{({\rm pole})}$
verifies both the gauge-fixing equation and the ghost equation, namely:
$$
{\delta \Gamma^{({\rm pole})}\over\delta B}=0,\quad
{\delta \Gamma^{({\rm pole})}\over\delta \bar c}-
\partial_{\mu}{\delta \Gamma^{({\rm pole})}\over\delta J_{\mu}}=0.
$$ 
Hence, so does the MS renormalized 1PI functional up to order $\hbar$:
$$
\eqalign{&{\delta \Gamma^{\rm MS}_{\ren}\over\delta B}=
\alpha B - \partial A\,+\,O(\hbar^2),\cr
&{\delta \Gamma^{\rm MS}_{\ren}\over\delta \bar c}-
\partial_{\mu}{\delta \Gamma^{\rm MS}_{\ren}\over\delta J_{\mu}}=0
\,+\,O(\hbar^2).\cr
}
$$

\section{Acknowledgments} We heartly thank T. Krajewski and R. Wulkenhaar for
pointing out to us a missing complex $i$ in our field strength and its 
bearing on the sign of the beta function; we corrected thus the wrong
sign ocurring in a previous computation of the beta function. We also
thank them for letting us know that they had shown
renormalizability and asymptotic freedom of the model estudied here on the
noncommutative torus \cite{\KW}.

\section{References}

\frenchspacing
\refno\Doplicher.
S. Doplicher, K. Fredenhagen and J.E. Roberts, Phys. Letts. B331 (1994) 39;
S. Doplicher, K. Fredenhagen and J.E. Roberts, Commun. Math. Phys. 172 (1995)
187.

\refno\CDS.
A. Connes, M.R. Douglas and A. Schwarz, J. High Energy. Phys. 02 (1998) 003;
{\tt hep-th/9711162}

\refno\Mtheory.
M.R. Douglas and
C. Hull, J. High Energy Phys. 9802 (1998) 008; M. Li, {\tt hep-th/9802052};
R. Casalbuoni, Phys. Lett. B431 (1998)69;
A. Schwarz, Nucl. Phys. B534 (1998)720;
T. Kawano and K. Okuyama, Phys. Lett. B433 (1998)29; 
Y.-K Cheung and M. Krog, Nucl. Phys. B528(1998)185
F. Ardalan, H. Arfaei 
and M.M. Sheikh-Jabbari, {\tt hep-th/9803067}; B. Moriaru and B. Zumino,
{\tt hep-th/9807198}; C. Hofman and E. Verlinde, {\tt hep-th/9810219}.

\refno\KLM.
P. Kosi\'nski, J. Lukierski and P. Ma\'slanka, {\tt hep-th/9902037}.

\refno\Madore.
J. Madore, {\it An Introduction to Noncommutative Geometry and its 
Applications}, LMS Lecture Notes 206 (1995).
 
\refno\Grosse.
H. Grosse, C. Klim{\v c}{\' i}k and P. Presnajder, Int. Jour. Theor. Phys.
35 (1996) 231; Commun. Math. Phys. 178 (1996)507, 180 (1996)429.

\refno\Filk.
T. Filk, Phys. Lett. B376 (1996)53.

\refno\VarPP.
J.C. V\'{a}rilly and J.M. Grac\'{\i}a-Bondia, {\tt hep-th/9804001}.

\refno\Chaichian.
M. Chaichian, A. Demichev and P. Pre{\v s}najder, {\tt hep-th/9812180}.

\refno\Instant.
N. Nekrasov and A. Schwarz, Commun. Math. Phys. 198 (1998) 689.

\refno\Book.
A. Connes, {\it Noncommutative Geometry}, Academic Press, 1994.

\refno\Landi.
G. Landi, {\it An Introduction to Noncommutative Spaces and their Geometries},
Springer Lecture Notes in Physics 51, Springer Verlag 1997.

\refno\Sorella.
O. Piguet and S.P. Sorella, {\it Algebraic Renormalization}, Springer-Verlag
1995.

\refno\Breitenlohner.
P.Breitenlohner and D. Maison, Commum. Math. Phys. 52 (1977) 11.

\refno\Speer.
E.R. Speer, {\it Generalized Feynman Amplitudes}, Princeton University Press
1969.

\refno\KW.
T. Krajewski and R. Wulkenhaar, {\it (Perturbative) Quantum Field Theory
on the Noncommutative Torus}, Leipzig seminar notes.

\bye